\documentclass[review]{elsarticle}

\usepackage[utf8]{inputenc}
\usepackage[T1]{fontenc}
\usepackage[english]{babel}

\usepackage[usenames,dvipsnames,svgnames,x11names]{xcolor}
\usepackage{todonotes}
\usepackage{amsmath}
\usepackage{csquotes}
\usepackage{graphicx}
\usepackage{siunitx}
\usepackage{pgfplots}
    \pgfplotsset{compat = 1.14}

\usetikzlibrary{patterns}

\usepackage{listings}
\def\lst{\lstinline}

\lstdefinelanguage{Rust}{%
  sensitive%
, morecomment=[l]{//}%
, morecomment=[s]{/*}{*/}%
, moredelim=[s][{\itshape\color[rgb]{0,0,0.75}}]{\#[}{]}%
, morestring=[b]{"}%
, alsodigit={}%
, alsoother={}%
, alsoletter={!}%
, morekeywords={break, continue, else, for, if, in, loop, match, return, with, while}  % control flow keywords
, morekeywords={as, const, let, move, mut, ref, static}  % in the context of variables
, morekeywords={dyn, enum, fn, impl, Self, self, struct, trait, type, union, use, where}  % in the context of declarations
, morekeywords={crate, extern, mod, pub, super}  % in the context of modularisation
, morekeywords={unsafe}  % markers
, morekeywords={abstract, alignof, become, box, do, final, macro, offsetof, override, priv, proc, pure, sizeof, typeof, unsized, virtual, yield}  % reserved identifiers
, morekeywords=[2]{Send}  % additional traits
, morekeywords=[3]{bool, char, f32, f64, i8, i16, i32, i64, isize, str, u8, u16, u32, u64, unit, usize, i128, u128}  % primitive types
, morekeywords=[4]{Err, false, None, Ok, Some, true}  % prelude value constructors
, morekeywords=[5]{assert!, assert_eq!, assert_ne!, cfg!, column!, compile_error!, concat!, concat_idents!, debug_assert!, debug_assert_eq!, debug_assert_ne!, env!, eprint!, eprintln!, file!, format!, format_args!, include!, include_bytes!, include_str!, line!, module_path!, option_env!, panic!, print!, println!, select!, stringify!, thread_local!, try!, unimplemented!, unreachable!, vec!, write!, writeln!}  % prelude macros
}%

\usepackage[printonlyused]{acronym}
    \newacro{AoS}{array of structures}
\newacro{IR}{intermediate representation}
\newacro{ISA}{instruction set architecture}
\newacro{MD}{molecular dynamics}
\newacro{MPI}{message passing interface}
\newacro{SIMD}{single instruction, multiple data}
\newacro{SoA}{structure of arrays}
\newacro{DEM}{discrete element methods}
\newacro{PBC}{periodic boundary conditions}

\usepackage[colorlinks=true,citecolor=blue,filecolor=blue,linkcolor=red,urlcolor=blue]{hyperref}
\usepackage{listings}

\addto\extrasenglish{%
}

\journal{Journal of Computational Science}

\begin{document}

\begin{frontmatter}

\title{tinyMD: A Portable and Scalable Implementation for Pairwise Interactions Simulations}
\tnotetext[mytitlenote]{This work is supported by the Federal Ministry of Education and Research (BMBF) as part of the MetaDL, Metacca, and ProThOS projects.}

\author{Rafael Ravedutti L.~Machado\fnref{lss}}
\author{Jonas Schmitt\fnref{lss}}
\author{Sebastian Eibl\fnref{lss}}
\author{Jan Eitzinger\fnref{rrze}}
\author{Roland Leißa\fnref{sic}}
\author{Sebastian Hack\fnref{sic}}
\author{Arsène Pérard-Gayot\fnref{sic}}
\author{Richard Membarth\fnref{sic,dfki}}
\author{Harald Köstler\fnref{lss}}

\fntext[lss]{Chair for System Simulation, University of Erlangen-Nürnberg.}
\fntext[rrze]{Regional Computer Center Erlangen (RRZE), University of Erlangen-Nürnberg.}
\fntext[sic]{Saarland University, Saarland Informatics Campus.}
\fntext[dfki]{German Research Center for Artificial Intelligence (DFKI), Saarland Informatics Campus.}




\begin{abstract}
% Molecular dynamics (MD) simulations are used to study the interactions among atoms and how these interactions affect their motion.
% These simulations are typically very compute-intensive and their performance is severely tied to the target architecture.
% Hence, it is important to optimize them for different hardware.
% Furthermore, distributing the computation over multiple processors or accelerator nodes requires to partition the simulation.
% In this case, it is important that the simulation scales well.
% But this is very challenging since communication between the compute nodes often harms the performance considerably.

% Abstract must contains 100 words at max
This paper investigates the suitability of the AnyDSL partial evaluation framework to implement tinyMD: an efficient, scalable, and portable simulation of pairwise interactions among particles.
We compare tinyMD with the \mbox{miniMD} proxy application that scales very well on parallel supercomputers.
We discuss the differences between both implementations and contrast miniMD's performance for single-node CPU and GPU targets, as well as its scalability on SuperMUC-NG and Piz~Daint supercomputers.
Additionaly, we demonstrate tinyMD's flexibility by coupling it with the waLBerla multi-physics framework. This allow us to execute tinyMD simulations using the load-balancing mechanism implemented in waLBerla.
\end{abstract}


\end{frontmatter}

%\linenumbers

\section{Introduction}

Nowadays, compute-heavy simulation software typically runs on different types of high-end processors to solve these simulations in a reasonable amount of time.
Nevertheless, this high-end hardware requires highly specialized code that is precisely adapted to the respective hardware in order to get anywhere near peak performance. 
What is more, in some cases different algorithmic variants are more suitable towards different kinds of hardware. 
For example, writing applications for a GPU, a massively parallel device with a very peculiar memory hierarchy, is very different from implementing applications for a CPU.
%for \ac{SIMD} parallelism,  GPU devices do not only require critical code to be implemented in a \ac{SIMD}-fashion, they also present a very distinct memory hierarchy in relation to general purpose processors.

Very large problems even call for distributed systems in which we have to partition the workload among multiple computers within a network.
This entails proper data communication between these systems;
transfer latencies should be hidden as much as possible.

%Besides the difference among target processors, we may also want to execute an application in distributed memory systems, where multiple processors split the work in order to perform it faster. This requires domain partitioning and data communication among different processes, these routines must properly distribute the workload (i.e.~through load balancing algorithms) and cannot consume a significant fraction of the execution time to maintain good parallel efficiency on these distributed systems.

In this paper we focus on \ac{MD} simulations.
These study the interactions among particles and how these interactions affect their motion.
To achieve peak performance on these simulations, the implementation must consider the best data access pattern for the target architecture.

We base our implementation \emph{tinyMD} on AnyDSL---a partial evaluation framework to write high-performance applications and libraries.
We compare this implementation with miniMD, a parallel and scalable proxy application that also contains GPU support;
miniMD is written in C++ and is based on Kokkos~\cite{6805038}.
Additionally, we also couple our tinyMD application with the waLBerla \cite{BAUER2020, godenschwager2013framework} multi-physics simulation framework. This allows us to exploit its load-balancing mechanism \cite{doi:10.1137/15M1035240} implementation within tinyMD simulations. We discuss the advantages that tinyMD provides in order to ease the coupling with different technologies.

We use the Lennard-Jones potential model in order to calculate the forces of the atoms in the experiments comparing to miniMD. Whereas in the experiments for load-balancing, we rely on the Spring-Dashpot force model, which is common in \ac{DEM} simulations~\cite{cundall79}.
Our goal is to compare and discuss the difference regarding the implementation and performance for both applications. We present experimental results for single-node performance in both CPU and GPU target processors, and we also show our experimental results for multi-node CPU processors in the SuperMUC-NG supercomputer, and multi-node GPU accelerators in the Piz~Daint supercomputer.

\subsection{Contributions}

In summary this paper makes the following contributions beyond our previous work~\cite{Schmitt18AnyDSL}:
\begin{itemize}
    \item We present our new tinyMD distributed-memory parallel implementation based upon AnyDSL and discuss its differences to miniMD---a typical C++ implementation based upon the Kokkos library to target GPU devices.
    For example, we use higher-order functions to build array abstractions.
    Due to AnyDSL's partial evaluator these abstractions are not accompanied by any overhead (see \autoref{sec:impl}).
    \item We demonstrate how flexible tinyMD is implemented with AnyDSL, and how its communication code can be coupled with the waLBerla framework to use its load-balancing feature in tinyMD simulations (see \autoref{sec:coupling}).
    \item We show performance and scalability results for various CPU and GPU architectures including multi-CPU results on up to 2048 nodes of the SuperMUC-NG cluster (98304 cores), and multi-GPU results on up to 1024 nodes of the Piz~Daint cluster (see \autoref{sec:eval}).
\end{itemize}
In order to make this paper as self-contained as possible, \autoref{sec:background} provides necessary background for both AnyDSL and \ac{MD} simulations after discussing related work in \autoref{sec:relwork}.

\section{Related Work}
\label{sec:relwork}

% miniMD, GROMACS, MESA-PD, LAMMPS, AMBER, NAMD, CHARMM, Desmond, https://arxiv.org/pdf/1704.00032.pdf

There is a wide effort on porting \ac{MD} simulations to different target architectures while delivering good performance and scalability. The majority of the developed frameworks and applications use the traditional approach of using a general-purpose language to implement the code for the simulations. % TODO forward reference

GROMACS~\cite{DBLP:journals/jcc/SpoelLHGMB05, ABRAHAM201519, pall2015} is a versatile \ac{MD} package used primarily for dynamical simulations of bio-molecules. It is implemented in C/C++ and supports most commonly used CPUs and GPUs. The package was initially released in 1991 and has been carefully optimized since then. GROMACS supports \ac{SIMD} in order to enhance the instruction-level parallelism and hence increase CPU throughput by treating elements as clusters of particles instead of individual particles \cite{PALL20132641}. GROMACS provides various hand-written \ac{SIMD} kernels for different \ac{SIMD} \acp{ISA}.

% https://github.com/gromacs/gromacs/tree/master/src/gromacs/simd
% TODO there is a huge list of GROMCAS-related publications.
% We should include at least some of them here.

LAMMPS~\cite{PLIMPTON19951, BROWN2012449} is an \ac{MD} code with focus on material modeling.
It contains potentials for soft and solid-state materials, as well as coarse-grain systems.
It is implemented in C++ and uses \ac{MPI} for communication when executed on several nodes.
% TODO there is a huge list of LAMMPS-related publications.
% We should include at least some of them here.

The Mantevo proxy application miniMD~\cite{DBLP:conf/pgas/LiLLHTP14,DBLP:journals/bioinformatics/RieberM17} is based on LAMMPS and performs and scales well (in a weak sense) on distributed memory systems, albeit with a very restricted set of features.
Since miniMD provides a very useful benchmark case for \ac{MD} simulations in terms of performance and scalability, we choose it as the base line for tinyMD.

MESA-PD~\cite{Eibl2019a} is a general particle dynamics framework.
Its design principle of separation of code and data allows to introduce arbitrary interaction models with ease.
It can thus also be used for molecular dynamics. Via its code generation approach using Jinja templates it can be adapted very effectively to any simulation scenario.
As successor of the pe rigid particle dynamics framework it inherits its scalability~\cite{Eibl2018} and advanced load balancing functionalities~\cite{Eibl2019}.

All these \ac{MD} applications have dedicated, hand-tuned codes for each supported target architecture.
This is the most common approach for today's \ac{MD} applications~\cite{doi:10.1002/wcms.1121,doi:10.5167/uzh-19245,doi:10.1002/jcc.20289}.
It requires much more effort as all these different code variants must not only be implemented, but also optimized, tested, debugged, and maintained independently from each other.

Other applications rely on domain-specific languages to generate parallel particle methods \cite{10.1145/3175659}, but this approach requires the development of specific compilation tools that are able to generate efficient code.
In this paper we explore the benefits from using the AnyDSL framework, where we shallow-embed \cite{10.1145/2692915.2628138, leissa2015} our domain-specific library into its front-end Impala and can then abstract device, memory layout and communication pattern through higher-order functions.
Thus, we use the compiler tool-chain provided by AnyDSL and do not need to develop specific compilation tools for \ac{MD} or particle methods.

\section{Background}
\label{sec:background}

\subsection{AnyDSL}
\label{sec:anydsl}

AnyDSL~\cite{DBLP:journals/pacmpl/LeissaBHPMSMS18} is a compiler framework designed to speed up the development of domain-specific libraries.
It consists of three major components:
the frontend \emph{Impala}, 
its \ac{IR} \emph{Thorin}~\cite{DBLP:conf/cgo/LeissaKH15}, 
and a runtime system.
The syntax of Impala is inspired from Rust and allows both imperative and functional programming.

\subsubsection{Partial Evaluation}

In contrast to Rust, Impala features a partial evaluator, which is controlled via \emph{filter expressions}~\cite{DBLP:conf/esop/Consel88}:
\begin{lstlisting}
fn @(?n) pow(x: i32, n: i32) -> i32 {
  if n == 1 {
    z
  } if n % 2 == 0 {
    let y = pow(x, n / 2);
    y * y
  } else {
    x * pow(x, n - 1)
  }
}
\end{lstlisting}
In this example, the function \lst{pow} is assigned the filter expression \mbox{\lst{?n}} (introduced via~\lst{@}).
%This expression is simply the operator \lst{?} applied to the parameter \lst{n}.
The \lst{?}-operator evaluates to \lst{true} whenever its argument is statically known by the compiler.
Now, at every call site, the compiler will instantiate the callee's filter expression by substituting all parameters with the corresponding arguments of that call site. 
If the result of this substitution is \lst{true}, the function will get executed and any call sites within will receive the same treatment.

In the case of the function \lst{pow} above, this means that the following call site will be partially evaluated because the argument provided for \lst{n} is known at compile-time:
\begin{lstlisting}
let y = pow(z, 5);
\end{lstlisting}
The result will be recursively expanded to:
\begin{lstlisting}
let y = z * pow(z, 4);
\end{lstlisting}
Then:
\begin{lstlisting}
let z2 = pow(z, 2);
let z4 = z2 * z2;
let y = z * z4;
\end{lstlisting}
And finally:
\begin{lstlisting}
let z1 = z;
let z2 = z1 * z1;
let z4 = z2 * z2;
let y = z * z4;
\end{lstlisting}
Note how this expansion is performed \emph{symbolically}.
In contrast to C++ templates or \mbox{\lst[morekeywords=constexpr]|constexpr|essions}, the Impala compiler does not need to know the value of the parameter \lst{x} to execute the function \lst{pow}.

\subsubsection{Triggered Code Generation}

Another important feature of Impala is its ability to perform \emph{triggered code generation}.
Impala offers built-in functions to allow the programmer to execute on the GPU, vectorize, or parallelize a given function.
% todo: is parallelization/vectorization mentioned at all in the paper?
For instance, the syntax to trigger code generation for GPUs that support CUDA is as follows:
\begin{lstlisting}
let acc = cuda_accelerator(device_index);
let grid = (n, 1, 1);
let block = (64, 1, 1);
with work_item in acc.exec(grid, block) {
    let id = work_item.gidx();
    buffer(id) = pow(buffer(id), 5);
}
\end{lstlisting}
This snippet launches a CUDA kernel on the GPU with index \lst{device_index} on a 1D grid and a block size of $64\times1\times1$.

\subsection{Molecular Dynamics}

\ac{MD} simulations are widely used today to study the behavior of microscopic structures.
These simulations reproduce the interactions among atoms in these structures on a macroscopic scale while allowing us to observe the evolution in the system of particles in a time-scale that is simpler to analyze.

Different areas such as material science to study the evolution of the system for specific materials, chemistry to analyze the evolution of chemical processes, or biology to reproduce the behavior of certain proteins and bio-molecules resort to simulations of \ac{MD} systems.

A system to be simulated constitutes of a number of atoms, the initial state (such as the atoms' position or velocities), and the boundary conditions of the system.
Here we use \ac{PBC} in all directions, hence when particles cross the domain, they reappear on the opposite side with the same velocity. 
%These simulations are defined by establishing the number of atoms in a system, the initial state of the system (such as the atoms' positions or velocities), and the boundary conditions for the system.
Fundamentally, the evolution of \ac{MD} systems is computed by solving Newton's second law equation, also known as the equation of motion (\autoref{eq:newton}). Knowing the force, it is possible to track the positions and velocities of atoms by integrating the same equation.
\begin{equation}
    F = m \dot{v} = m a \label{eq:newton}
\end{equation}
The forces of each atom are based on its interaction with neighboring atoms.
This computation is usually described by a potential function or force field.
Many different potentials can be used to calculate the particle forces, and each one is suitable for different types of simulation.
In this work, the potential used for miniMD comparison is the Lennard-Jones potential (\autoref{eq:lennard_jones})---a pair potential for van-der-Waals forces.
Consider $x_i$ the position of the particle $i$, the force for the Lennard-Jones potential can be expressed as:
\begin{equation}
    F_{2}^{LJ}(x_i, x_j) = 24\epsilon \left( \frac{\sigma}{x_{ij}} \right)^{6} \left[ 2\left(\frac{\sigma}{x_{ij}}\right)^{6} - 1\right] \frac{x_{ij}}{|x_{ij}|^{2}}
    %V_{LJ} = 4\epsilon \left[ \left(\frac{\sigma}{r}\right)^{12} - \left(\frac{\sigma}{r}\right)^6 \right]
    \label{eq:lennard_jones}
\end{equation}
Here, $x_{ij}$ is the distance vector between the particles $i$ and $j$, $\epsilon$ determines the width of the potential well, and $\sigma$ specifies at which distance the potential is~$0$.

For our load balancing experiments, we use the Spring-Dashpot (\autoref{eq:spring_dashpot}) contact model to provide a static simulation and to demonstrate the flexibility of our application. This contact model is commonly used on \ac{DEM} to simulate rigid bodies interactions (instead of point masses on MD simulations). Therefore we consider particles as spheres on these experiments. Consider the position $x_i$ and velocity $v_i$ for the particle $i$. Its force is defined as:
\begin{align}
    F_{2}^{SD}(x_i, v_i, x_j, v_j) &= K\xi + \gamma\dot{\xi}
    \label{eq:spring_dashpot} \\
\intertext{where}
    \xi       &=  \hat{x}_{ij}(\sigma - |x_{ij}|)\Theta(\sigma - |x_{ij}|) \\
    \dot{\xi} &= -\hat{x}_{ij}(\hat{x}_{ij} \cdot v_{ij})\Theta(\sigma - |x_{ij}|)
\end{align}
and $K$ being the spring constant, $\gamma$ the dissipation constant, $\hat{x}_{ij}$ the unit vector $\frac{x_{ij}}{|x_{ij}|}$, $\sigma$ the sphere diameter, and $\Theta$ the Heaviside step function.

Naively, an \ac{MD} implementation iterates over all atom pairs in the system.
This requires to consider a quadratic number of pairs.
For short range interactions, however, we can use a Verlet list~\cite{PhysRev.159.98} instead to keep track of all nearby particles within a certain \emph{cutoff radius}.
Then, we compute the potential force for an atom only for the neighbors stored in its list.
We regularly update the Verlet list to keep track of new atoms that enter the cutoff radius.

\begin{figure}[t]
\includegraphics[width=4cm]{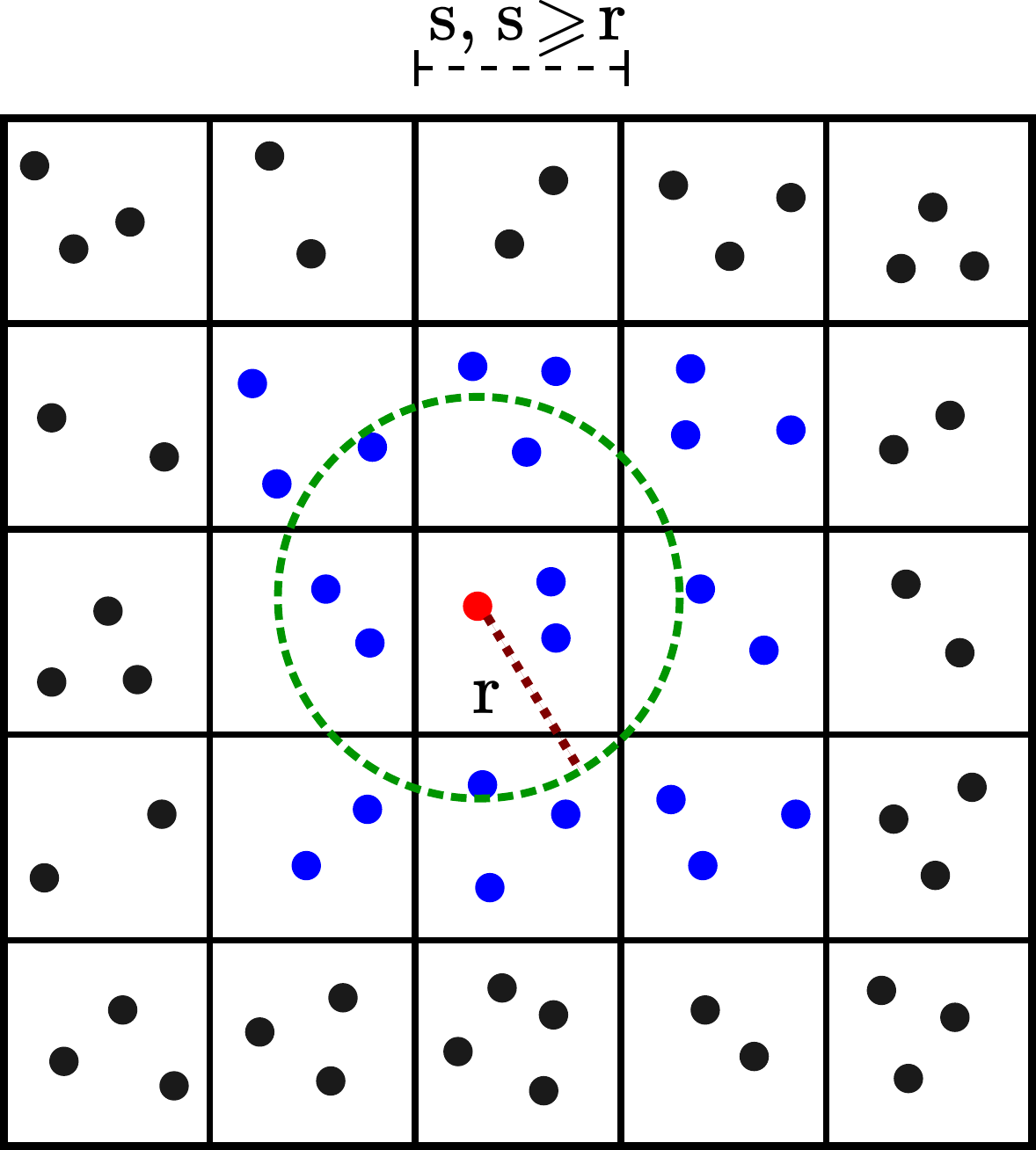}
\centering
\caption{Neighbor list creation example. In this case the neighbor list is built for the red particle, and only the blue particles in the neighbor cells are checked for comparison. Particles within the green area with radius $r$ are inserted into the red particle neighbor list. The size for the cells $s$ can be greater or equal than $r$, but not less than it\protect\footnotemark. The value for $r$ is usually the cutoff radius plus a small value (called verlet buffer) so neighbor lists do not have to be updated every time step.}
\label{fig:neighborlists}
\end{figure}

Although it is not necessary to build the Verlet list at all time steps, it is still a costly procedure since it requires the iteration over each pair of atoms.
To enhance the building performance, cell lists can be used to segregate atoms according to their spatial position.
As long as the cell sizes are equal to or greater than the cutoff radius, it is just necessary to iterate over the neighbor cells to build the Verlet list.
\autoref{fig:neighborlists} depicts the creation of the neighbor lists for a particle using cell lists.
Neighbor lists can be created for just half of the particle pairs (known as half neighbor lists).
This allows for simultaneous updates of both particle forces within the same iteration (the computed force is subtracted from the neighbor particle) but requires atomic operations to prevent race conditions.

\footnotetext{In some implementations the cell size can be less than the cutoff radius, however the cell neighborhood to be checked must be extended accordingly.}

% In order to allow for \ac{SIMD} vectorization, it is beneficial to cluster particles as opposed to treating them individually.
% The cluster size corresponds to the desired vector length.
% This enables us to perform all calculations on a whole cluster of particles in \ac{SIMD} manner.
% In addition, storing these clusters contiguously in memory allows for efficient vector loads and stores.
% For a fair comparison, we just use the Verlet list and cell list strategies in tinyMD since miniMD cannot cluster particles.

\section{The tinyMD Library}
\label{sec:impl}

In this section we introduce and discuss tinyMD\footnote{\url{https://github.com/AnyDSL/molecular-dynamics}}.
We focus on the main differences in writing portable code with AnyDSL as opposed to traditional C/C++ implementations.
We explore the benefits achieved by using higher-order functions to map code to different target devices, different data layouts and to implement flexible code for \ac{MPI} communication.
In the following we use the term \emph{particle} to refer to atoms.% since it is the term employed in tinyMD.

\subsection{Device Mapping}
\label{sec:device_mapping}

In order to map code parts for execution on the target device,  tinyMD relies on the \lst{Device} abstraction.
It contains functions to allocate memory, transfer data, launch a loop on the target device, and perform even more complex device-dependent procedures such as reductions:
\begin{lstlisting}
struct Device {
  alloc: fn(i64) -> Buffer,
  transfer: fn(Buffer, Buffer) -> (),
  loop_1d: fn(i32, fn(i32) -> ()) -> (),
  // ...
}
\end{lstlisting}
Similar to a Java interface or abstract virtual methods in C++, a \mbox{\lst|Device|} instance allows tinyMD to abstract from the concrete implementation of these functions---in this case device-specific code.
Unlike Java interfaces or virtual methods however, the partial evaluator will remove these indirections by specializing the appropriate device-specific code into the call-sites.
Each device supported in tinyMD possesses its own implementation: a function that returns a \lst|struct| instance which contains several functions, and hence, \enquote{implements the \lst|Device| interface}.
For example, the CPU implementation looks like this:
\begin{lstlisting}
fn @device() -> Device {
  Device {
    alloc: |size| { alloc_cpu(size) },
    transfer: |from, to| {}, // no copy required
    loop_1d: @|n, f| {
      vectorized_range(get_vector_width(), 0, n, |i, _| f(i));
    },
    // ...
  }
}
\end{lstlisting}
The \lst{vectorized_range} function iterates over the particles. 
This function in turn calls the \lst{vectorize} intrinsic that triggers the \emph{Region Vectorizer}~\cite{10.1145/3296979.3192413} to vectorize the LLVM \ac{IR} generated by the Impala compiler. Although not covered in this paper, tinyMD employs the \lst|parallel| intrinsic to spawn multiple threads for multiple CPU cores.

The GPU implementation looks like this:
\begin{lstlisting}
fn @device() -> Device {
  Device {
    alloc: |size| { acc.alloc(size) },
    transfer: |from, to| { copy(from, to) },
    loop_1d: @|n, f| {
      // build grid_size and block_size for n
      acc.exec(grid_size, block_size, |work_item| {
        let i = work_item.bidx() * work_item.bdimx() + work_item.tidx();
        if i < n { f(i); }
      });
    },
    // ...
  }
}
\end{lstlisting}

The \lst|Device| abstraction is sufficient to map our application to different targets, it takes care of separating the parallel execution strategy from the compute kernels and provides basic functions for the device. Together with the data management abstractions (see \autoref{sec:data}) we attain performance portability with substantially less effort.

\subsection{Data Management}
\label{sec:data}

To store our simulation data, we built an \lst{ArrayData} structure, which defines a multi-dimensional array in tinyMD.
This data structure takes care of memory allocation in both device and host (if required).
The actual accesses are abstracted from with the help of \lst|ArrayLayout|.
This idiom is similar to the \lst|Device| abstraction and makes arrays in tinyMD both target and layout-agnostic. 
%and can be bound to an \lst{ArrayLayout} data structure that provides abstractions for the data layout:
\begin{lstlisting}
struct ArrayData {
  buffer: Buffer,
  buffer_host: Buffer,
  size_x: i32,
  size_y: i32,
  host_mirror: bool
};

struct ArrayLayout {
  index_2d_fn: fn(ArrayData, i32, i32) -> i32,
  add_fn: fn(ArrayData, i32, real_t) -> ()
};
\end{lstlisting}
With this definition, \lst{ArrayData} just contains sizes for the \lst|x| and \lst|y| dimensions. This can be easily extended to more dimensions, but for our application, two dimensions are enough. The following list demonstrates how to implement a simple row-major order array layout similar to C arrays (again, this is akin to \enquote{implementing the \lst|ArrayLayout| interface}):
\begin{lstlisting}
fn @row_major_order_array(is_atomic: bool) -> ArrayLayout {
  ArrayLayout {
    index_2d_fn: @|array, x, y| { x * array.size_y + y },
    add_fn: @|array, i, v| { do_add(is_atomic, array, i, v) }
  }
}
\end{lstlisting}
Note that the layout can also be made atomic, by using the \lst{do_add} function that uses atomic addition when the first argument \lst{is_atomic} is \lst{true}. The AnyDSL partial evaluator takes care of generating the proper add instructions in the final code with zero overhead when the \lst{is_atomic} flag is known at compile-time. Similarly, we define a column-major order layout as used in Fortran:
%\todo{maybe show an implementation of add\_fn that works on the address of the array}
\begin{lstlisting}
fn @column_major_order_array(is_atomic: bool) -> ArrayLayout {
  ArrayLayout {
    index_2d_fn: @|array, x, y| { y * array.size_x + x },
    add_fn: @|array, i, v| { do_add(is_atomic, array, i, v) }
  }
}
\end{lstlisting}

For a more complex data layout, we show the definition of a clustered array, also known as an Array of Struct of Arrays (AoSoA). This is an array of structs whose elements are again clusters of elements as opposed to individual elements. With cluster sizes that are a power of two, we get:
\begin{lstlisting}
fn @clustered_array(is_atomic: bool, cluster_size: i32) -> ArrayLayout {
  let mask = cluster_size - 1;
  let shift = get_shift_size(cluster_size) - 1;
  ArrayLayout {
    index_2d_fn: @|array, x, y| {
      let i = x >> shift;
      let j = x & mask;
      cluster_size * (i * array.size_y + y) + j
    },
    add_fn: @|array, i, v| { do_add(is_atomic, array, i, v) }
  }
}
\end{lstlisting}

For most common use cases, the cluster size is known at compile-time, which allows AnyDSL to specialize this data layout by directly replacing the pre-computed values of \lst{shift} and \lst{mask}. If more than one cluster size is used in the application, different specialized codes are generated.

To bind both array structures mentioned above, we write different template functions that perform operations such as \lst{get} and \lst{set} on values. We also provide target functions to abstract whether the device or the host buffer must be used. The following snippet shows these target functions and the template for reading \lst{real_t} elements in a 2-dimensional array:

\begin{lstlisting}
fn @array_dev(array: ArrayData) -> Buffer { array.buffer }
fn @array_host(array: ArrayData) -> Buffer { array.buffer_host }
type ArrayTargetFn = fn(ArrayData) -> Buffer;

fn @array_2d_get_real(
  target_fn: ArrayTargetFn, layout: ArrayLayout, array: ArrayData,
  i: i32, j: i32) -> real_t {

  bitcast[&[real_t]](target_fn(array).data)(layout.index_2d_fn(array, i, j))
}
\end{lstlisting}

In this case, we also resort to the partial evaluator to generate the specialized functions for all used targets and layouts. We can also map non-primitive data types to our arrays, the following example shows how to map our \lst{Vector3D} structure to N$\times3$ arrays. The structure contains \lst{x}, \lst{y} and \lst{z} elements of type \lst{real_t}:

\begin{lstlisting}
// Get Vector3D value in 2D array with abstract target and layout
fn @array_2d_get_vec3(
    target_fn: ArrayTargetFn, layout: ArrayLayout, array: ArrayData,
    i: i32) -> Vector3D {

  Vector3D {
    x: array_2d_get_real(target_fn, layout, array, i, 0),
    y: array_2d_get_real(target_fn, layout, array, i, 1),
    z: array_2d_get_real(target_fn, layout, array, i, 2)
  }
}
\end{lstlisting}

Notice that through \lst{set()} and \lst{get()} functions it is also possible to abstract data over scratchpad memory (such as shared or texture memory on GPU devices). This could be done by first staging data into these memory and then providing set and get functions that operate on them.

Finally, we can use these templates functions to implement the abstractions for our particle data. For this, we created a \lst{Particle} data structure that holds the proper functions to write and read particle information:

\begin{lstlisting}
struct Particle {
  set_position: fn(i32, Vector3D) -> (),
  get_position: fn(i32) -> Vector3D,
  // ...
};
\end{lstlisting}

The \lst{Particle} interface takes care of generating easy to use functions that map particle information to our \lst{ArrayData} primitives. These functions can set and get particle properties, iterate over the neighbor lists and perform any operation that relies on the data layout and target. Thus, we can generate \lst{Particle} structures for specific targets (host or device) and for distinct layouts. This is done through the \lst{make_particle} function:

\begin{lstlisting}
fn @make_particle(
    grid: Grid, target_fn: ArrayTargetFn,
    vec3_layout: ArrayLayout, nb_layout: ArrayLayout) -> Particle {

  Particle {
    set_position: @|i, p| array_2d_set_vec3(target_fn, vec3_layout, grid.positions, i, p),
    get_position: @|i|    array_2d_get_vec3(target_fn, vec3_layout, grid.positions, i),
    //...
  }
}
\end{lstlisting}

The structure generated by \lst{make_particle} contains the specialized functions for the layouts we specified, these functions access buffers in the memory space provided by the target function. To achieve a full abstraction, we then put it altogether with the device loop and the layout definition:

\begin{lstlisting}
fn @ParticleVec3Layout() -> ArrayLayout { row_major_order_array(false) }
fn @particles(grid: Grid, f: fn(i32, Particle) -> ()) -> () {
  device().loop_1d(grid.nparticles, |i| {
    f(i, make_particle(grid, array_dev, ParticleVec3Layout(), NeighborlistLayout()));
  });
}
\end{lstlisting}

Data layout definitions such as the \lst{ParticleVec3Layout} can also be written in the device-specific codes.
Consider the \lst{NeighborlistLayout} layout as an example: For the CPU it is optimal to store data in a particle-major order to enhance locality per thread during access, and therefore improving cache utilization.
For GPUs, neighbor-major order is preferable because it enhances coalesced memory access.
%\todo{no implementation for NeighborlistLayout is shown, is this intended?}
Since each GPU thread computes a different particle, we keep all the nth neighbor for each particle contiguous in memory, causing threads to access these data in the same iteration and therefore reducing the number of required transactions to load the entire data.
Nevertheless, the \lst{NeighborlistLayout} has the same definition as the \lst{ParticleVec3Layout} shown above for CPU targets, and is defined as a \lst{column_major_order_array} for GPU targets.

To show how simple it is to use our \lst{particles} abstraction, consider the following example to compute particle-neighbor potentials:
%\todo{Here it would be nice to show how i, j and pos\_i and pos\_j are used. should pos\_i be passed to particle.neighbors?}
\begin{lstlisting}
particles(grid, |i, particle| {
  let pos_i = particle.get_position(i);
  particle.neighbors(i, |j| {
    let pos_j = particle.get_position(j);
    let del = vector_sub(pos_i, pos_j);
    let rsq = vector_len2(del);
    if rsq < rsq_cutoff {
      let f = potential(del, rsq);
      // update force for i (and j if half neighbor lists is being
      // used) with the computed force
    }
  });
});
\end{lstlisting}

We also provide a \lst{compute_potential} syntactic-sugar that can be used as follows to compute the Lennard-Jones potential (consider the lamba-function passed as the \lst{potential} function used previously):
%\todo{I guess we have to explain the signature of compute\_potential. Will a reader understand why a lambda-function is passed as last argument?}

\begin{lstlisting}
let sigma6 = pow(sigma, 6);
compute_potential(grid, half_nb, rsq_cutoff, @|del, rsq| {
  let sr2 = 1.0 / rsq;
  let sr6 = sr2 * sr2 * sr2 * sigma6;
  let f = 48.0 * sr6 * (sr6 - 0.5) * sr2 * epsilon;
  vector_scale(f, del) // returns the force to be added
});
\end{lstlisting}

AnyDSL can generate specialized variants for \lst{compute_potential} with full- and half-neighbor lists through the \lst{half_nb} parameter.
This avoids extra conditions on full neighbor lists kernels that are just required with half neighbor lists.
The same specialization is performed when building the neighbor lists.

All these abstractions provide a very simple and extensible way to work with different devices and data management.
The template functions and layout specifications can be extended to support more complex operations and map to different data types.
Furthermore, these can be used to yield a domain-specific interface (such as the \lst{Particle} abstraction) to improve the usability of our library.
This is accomplished with no overhead in the final generated code with AnyDSL.

% Although we do not have all the features that Kokkos has, our abstractions required just a few hundred lines of code, a much smaller effort compared to implement and maintain an entire framework.
%\todo{do we show LoC later on?}

\subsection{Communication}
\label{sec:comm}

The communication for our code can be separated in three routines as is also done in miniMD.
These routines are listed below:

\begin{itemize}
    \item Exchange: exchange particles that overlap the domain for the current rank, this particle becomes a local particle on the process it is sent to. This operation is not performed on every time step, but at an interval of $n$ time steps, $n$ is adjustable by the application.
    \item Border definition: defines the particles that are in the border of the current rank's domain, these particles are sent as ghost particles to the neighbor processes. This routine also is performed at every $n$ time steps.
    \item Synchronization: uses the border particles defined in the border definition, and just sends them at every time step of the simulation. Since the number of particles to be sent is known beforehand, it is a less costly routine.
\end{itemize}

We provide a generic way to abstract the communication pattern on these routines through a higher-order function named \lst{communication_ranks}.
This function receives both conditional functions to check whether particles must be exchanged or sent as ghost particles to neighbor ranks.
\ac{PBC} correction can also be proper applied using these conditional functions, which greatly improves the flexibility for our communication code.

These conditional functions also help when coupling tinyMD with waLBerla because they separate the logic to check particle conditions from the packing and MPI send/recv routines that stay untouched.
When using waLBerla domain partitioning, particle positions must be checked against the supplied waLBerla data structures (block forest domain), whereas for miniMD pattern we can just write simple comparisons in the 6 directions.
The \lst{communication_ranks} implementation separating both strategies is listed as follows:

\begin{lstlisting}
// Types for condition and communication functions
type CondFn = fn(Vector3D, fn(Vector3D) -> ()) -> ();
type CommFn = fn(i32, i32, CondFn, CondFn) -> ();
fn communication_ranks(grid: Grid, body: CommFn) -> () {
  if use_walberla() { // Use walberla for communication
    //...
    range(0, nranks as i32, |i| {
      let rank = get_neighborhood_rank(i);
      body(rank, rank, // Conditions to send and receive from rank
      @|pos, f| { /* Check for border condition with walberla */ },
      @|pos, f| { // Exchange condition
        for j in range(0, get_rank_number_of_aabbs(i)) {
          let p = pbc_correct(pos); // PBC correction
          if is_within_domain(p, get_neighbor_aabb(get_rank_offset(i) + j)) {
            f(p);
            break()
          }
        }
      });
    });
  } else { // Use 6D stencil communication like miniMD
    body(xnext, xprev, // Conditions to send to xnext
    @|p, f| { if p.x > aabb.xmax - spacing * 2.0 { f(pbc_correct(p)); }},
    @|p, f| { if p.x > aabb.xmax - spacing  { f(pbc_correct(p)); }});
    body(xprev, xnext, // Conditions to send to xprev
    @|p, f| { if p.x < aabb.xmin + spacing * 2.0 { f(pbc_correct(p)); }},
    @|p, f| { if p.x < aabb.xmin + spacing  { f(pbc_correct(p)); }});
    // Analogous for y and z dimensions
  }
}
\end{lstlisting}

The \lst{communication_ranks} function can be used to pack particles and to perform the MPI communications since it provides the functions to check for which particles must be packed and the ranks to send and receive data.
The following snippet shows a simple usage to obtain the particles to be exchanged:

\begin{lstlisting}
communication_ranks(grid, |rank, _, _, exchange_positions| {
  particles(grid, |i, particle| {
    exchange_positions(particle.get_position(i), @|pos| {
      // Here pos is a particle position that must be exchanged,
      // it already contains the proper PBC adjustments
    });
  });
});
\end{lstlisting}

On miniMD, a 6-stencil communication pattern is hard-coded to the simulation, which turns the code inflexible to be coupled to other technologies.
This also shows an important benefit from higher-order functions, as code can be more easily coupled with other technologies by simply replacing functionality.

\subsection{Summary of Benefits}

This section summarizes the benefits of using the AnyDSL framework for both tinyMD and \ac{MD} applications in general.

\paragraph{Separation of Concerns}
We separate our force potential computation logic from device-specific code and rely on higher-order functions to map it to the proper target.
This substantially reduces the effort in writing portable applications.
We employ the same technique to abstract other procedures such as the \ac{MPI} communication calls.
Here, we adopt higher-order functions to define communication patterns (see \autoref{sec:comm}).

\paragraph{Layers of Abstractions}
We hide device-dependent code like our data layout (see \autoref{sec:data}) or device mapping (see \autoref{sec:device_mapping}) behind functions.
These abstractions have zero overhead due to the partial evaluator (see below).

\paragraph{Compile-Time Code Specialization}
We utilize Impala's partial evaluator to generate faster specialized code when one or more parameters are known at compile-time.
This significantly reduces the amount and complexity of the code while still being able to generate all desired variants at compile time.

\section{Coupling tinyMD with waLBerla}
\label{sec:coupling}

\begin{figure}[t]
\includegraphics[width=12cm]{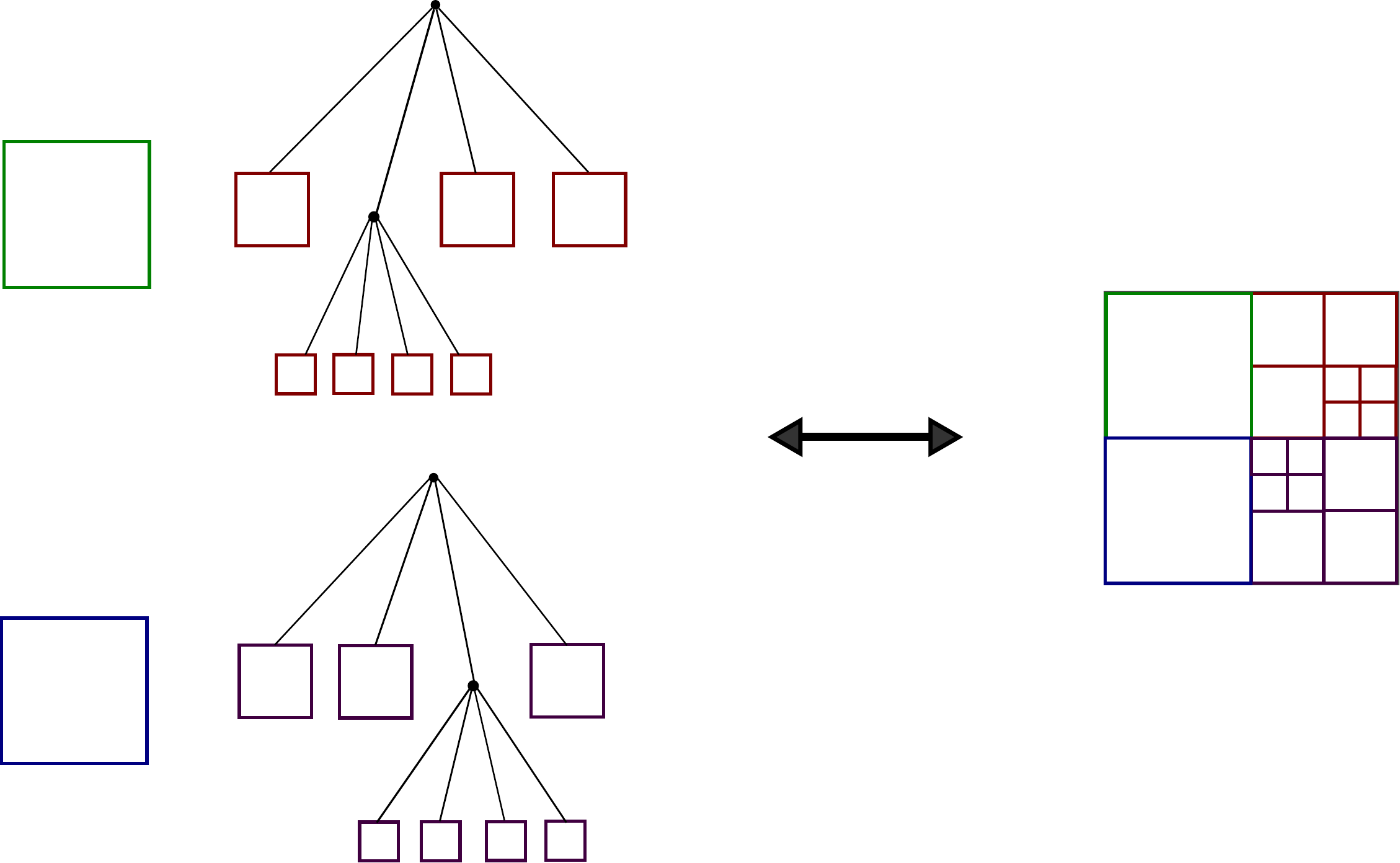}
\centering
\caption{Schematic 2D illustration of the waLBerla block forest domain partitioning. At left, the forest of octrees is depicted, where it is possible to observe the refinement of the blocks. At right, the domain partitioning for the correspondent block forest is depicted. On 3D simulations, refined blocks produce 8 child blocks in the tree instead of 4.}
\label{fig:block_forest}
\end{figure}

In this section, we briefly present the fundamental concepts behind waLBerla to understand its load balancing mechanism.
The most important characteristic is its domain partitioning using a forest of octrees called block forest. This kind of partitioning allows us to refine blocks in order to manage and distribute regions with smaller granularities.
Furthermore, we explain how this block forest feature written in C++ is integrated into our tinyMD Impala code.

waLBerla is a modern multi-physics simulation framework that supports massive parallelism of current peta- and future exascale supercomputers.
Domain partitioning in waLBerla is done through a forest of octrees. The leaf nodes, so called blocks, can be coarsened or refined and distributed among different processes. \autoref{fig:block_forest} depicts the waLBerla forest of octrees data structure with its corresponding domain partitioning.

For each local block, waLBerla keeps track of its neighbor blocks and the process rank that owns it.
This information is important to determine to which processes we must communicate and with which blocks/subdomains we need to compare our particle positions.

To balance the simulation such that work is evenly distributed among the processes, it is first necessary to assign weights to each block in the domain.
In this paper we use the number of particles located on a block as the weight of the block.
The weight is not only used during the distribution of blocks, it is also used to determine whether a block should be refined (when it reaches an upper threshold) or merged (when it reaches a lower threshold).
Different algorithms are available in waLBerla to distribute the workload. The algorithms can be categorized into space-filling curves \cite{bader2013space}, graph partitioning and diffusive schemes.
In this paper we concentrate on space-filling curves, more specifically the Hilbert \cite{Campbell2003} and Morton \cite{Morton1966} (or z-order) curves.

To couple applications and combine their functionality can save a lot of effort as one does not have to re-implement the functionality.
In this paper, we chose to couple the load balancing implementation from waLBerla with tinyMD. As this part of the code does not have to be portable to different devices we can rely on an external framework.
One could try to implement a portable code for it, but the amount of work and complexity to do so may not be worth the benefits.
Nevertheless, we can exploit AnyDSL to optimize our simulation kernels, memory management and other parts of the application that can be advantageous, and then rely on existing implementations to avoid redoing work that does not give benefits in the end.

Since our Impala implementation turns out to be compiled to LLVM IR at some point, we can link C/C++ code with it using Clang.
Therefore, we write an interface from Impala and C++ in order to do the coupling. \autoref{sec:comm} already presented how our communication code is integrated through the usage of routines that fetch information provided by waLBerla.
In this section, we focus on how these routines and other parts of the load balancing are written.

The first part is to initialize the waLBerla data structures, the block forest is created using the bounding box information from tinyMD.
When the block forest for a process is adjusted due to the balancing, the tinyMD bounding box is also changed, and since tinyMD does not have a block forest implementation, we just transform the waLBerla block forest into a simple AABB structure.
This is done by performing an union of all blocks that belongs to the current process.
In the end, the domain can occupy a larger region than necessary, but this does not affect the results, just the memory allocation.

\begin{figure}[t]
\includegraphics[width=12cm]{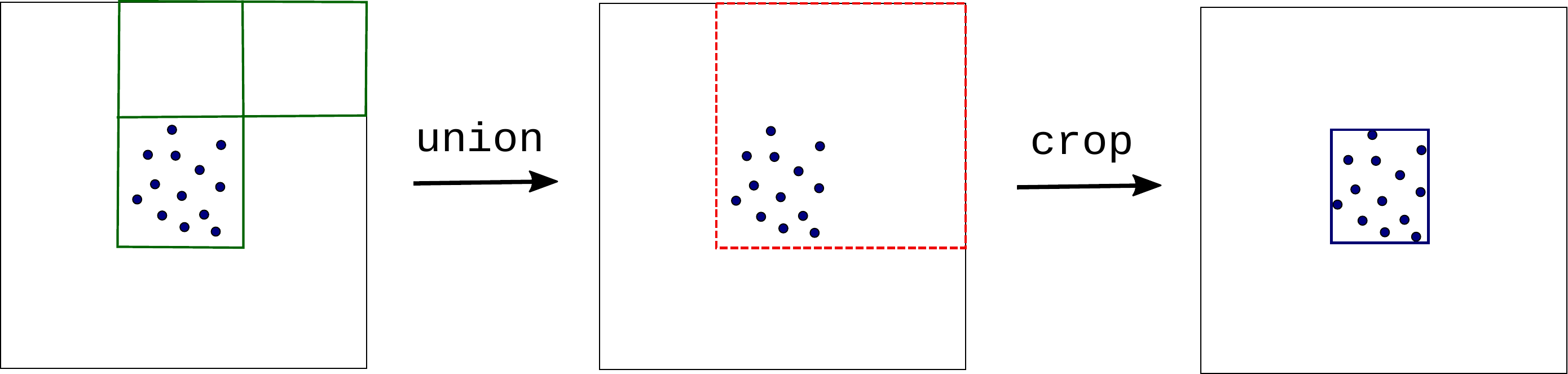}
\centering
\caption{Transformation from waLBerla block forest domain to tinyMD regular AABB. The union of blocks is performed so all blocks that belong to the current process are converted to tinyMD simple AABB structure, which only supports rectangular domains. Crop is performed to remove empty space and reduce the amount of allocated memory. On simulations which the empty space size is not significant, the crop operation can be skipped.}
\label{fig:aabb_transformation}
\end{figure}

In some extreme cases where we only fill part of the global domain with particles, the whole empty part of the domain can end up being assigned to a single process, which turns it to allocate an amount of memory proportional to the global domain.
To mitigate this issue, we simply define a process bounding box as enough to comprise its particles---or in other words, we crop the domain.
Since tinyMD bounding box is only useful to define cells and distribute the particles over them, this does not affect the correctness of the simulation.
\autoref{fig:aabb_transformation} depicts how the union and cropping transform the grid on tinyMD.

To crop the AABB, a reduction operation is required. The following code shows how this is performed using the \lst{reduce_aabb} function abstracted by the device:

\begin{lstlisting}
let b = AABB {
  xmin: grid.aabb.xmax,
  xmax: grid.aabb.xmin,
  // ... analogous to y and z
};

let red_aabb_fn = @|aabb1: AABB, aabb2: AABB| {
  AABB {
    xmin: select(aabb1.xmin < aabb2.xmin, aabb1.xmin, aabb2.xmin),
    xmax: select(aabb1.xmax > aabb2.xmax, aabb1.xmax, aabb2.xmax),
    // ... analogous to y and z
  }
};

let aabb = device().reduce_aabb(grid.nparticles, b, red_aabb_fn, @|i| {
  let pos = particle.get_position(i);
  AABB {
    xmin: pos.x,
    xmax: pos.x
    // ... analogous to y and z
  }
});
\end{lstlisting}

Note that \lst{reduce_aabb} is device-specific and is optimally implemented on GPU, which also demonstrates benefits obtained through the \lst{Device} abstraction presented on \autoref{sec:device_mapping}.
One can also observe how simple it is with AnyDSL to execute device code with non-primitive data types, in this case the \lst{AABB} structure.
Furthermore, reduction can also be used to count the number of particles within a domain, which is useful to obtain the weights for the load-balancing, the following code shows how it is written in tinyMD:

\begin{lstlisting}
let sum = @|a: i32, b: i32| { a + b };
computational_weight = device().reduce_i32(nparticles, 0, sum, |i| {
  select(is_within_domain(particle.get_position(i), aabb), 1, 0)
}) as u32;

communication_weight = device().reduce_i32(nghost, 0, sum, |i| {
  select(is_within_domain(particle.get_position(nparticles+i), aabb), 1, 0)
}) as u32;
\end{lstlisting}

Next step for the coupling is to provide a function in tinyMD to update the neighborhood of a process.
We first build the list of neighbors using the waLBerla API in C++ and then call a tinyMD function to update the neighborhood data in tinyMD.
An important part of the process is to perform a conversion from the dynamic C++ container data types to simple arrays of \lst{real_t} (in case of the boundaries) and integers (in case of neighbor processes ranks and number of blocks per neighbor process).
This conversion is necessary because (a) Impala does not support this dynamic types from C++ and (b) code that executes on GPU, more specifically the particle position checking presented on \autoref{sec:comm} also does not support these dynamic data types.

Finally, it is also necessary to perform serialization and de-serialization of our particle data during the balancing step.
This is required because the blocks data must be transferred to their new process owners during the distribution.
Since waLBerla is an extensible framework, it provides means to call custom procedures when a block must be moved from or to another process.
This allow us to implement (de)serialization functions on tinyMD, which also updates the local particles that are exchanged.
The communication in these routines is entirely handled by waLBerla, hence no MPI calls are performed by tinyMD.

\section{Evaluation}
\label{sec:eval}

We evaluated tinyMD as well as miniMD on several CPU and GPU architectures.
We chose the following CPUs:

\medskip
\begin{tabular}{l@{\phantom{X}}lr}
    \textbf{Cascade Lake:}  & Intel(R) Xeon(R) Gold 6246 CPU          & @ 3.30\,GHz \\
    \textbf{Skylake:}       & Intel(R) Xeon(R) Gold 6148 CPU          & @ 2.40\,GHz \\
    \textbf{Broadwell:}     & Intel(R) Xeon(R) CPU E5-2697 v4         & @ 2.30\,GHz
\end{tabular}
\medskip

\noindent
And the following GPUs:

\medskip
\begin{tabular}{l@{\phantom{X}}lr}
    \textbf{Pascal:} & GeForce GTX 1080     & ( 8\,GB memory) \\
    \textbf{Turing:} & GeForce RTX 2080 Ti  & (11\,GB memory) \\
    \textbf{Volta:}  & Tesla V100-PCIe-32GB & (32\,GB memory)
\end{tabular}
\medskip

We ran each simulation over 100 time steps---each time step with 0.005. %\todo{seconds?}
% we use the lj unit as miniMD/LAMMPS, so quantities are unit-less and based on epsilon and sigma parameters
% https://lammps.sandia.gov/doc/units.html
We performed particle distribution over cells and reneighboring every 20 time-steps with double-precision floating point and full neighbor interaction.
We use the Lennard-Jones potential with parameters $\epsilon = 1$ and $\sigma = 1$ (see \autoref{eq:lennard_jones}). The particles setup in tinyMD is the same as in miniMD, as well as the cutoff radius of 2.5 and the Verlet buffer of 0.3.

For multi-node benchmarks, particles are exchanged with neighbor ranks at each 20 time-steps before distribution over cells and reneighboring, and communications to update the particle positions within ghost layers are done every time-step with neighbor ranks.

For CPU, tests were performed in a single core (no parallelism) with fixed frequency, we simulated a system configuration of $32^3$ unit cells with 4~particles per unit cell.
For tinyMD, the Impala compiler generates Thorin code that is further compiled with Clang/LLVM~8.0.0.
For miniMD, we use the Intel Compiler (icc) 18.0.5.

\colorlet{force_color_cpu}{blue!70}
\colorlet{neigh_color_cpu}{blue!50}
\colorlet{other_color_cpu}{blue!30}

\begin{figure}[t]
\centering
\begin{tikzpicture}[scale=0.95]
    \pgfplotsset{
        ybar stacked, ymin=0, ymax=18, xmin=0.5, xmax=3.5, xtick=data,
        xtick={1,...,3},
        xticklabels={Broadwell, Cascade Lake, Skylake},
        xticklabel style={yshift=-10pt},
        ylabel={time to solution (s)}, ylabel style={yshift=-1ex},
        legend cell align={left},
        /pgf/bar width=8pt,% bar width
        scatter/position=absolute,
        node near coords style={
            font=\footnotesize,
            at={(axis cs:\pgfkeysvalueof{/data point/x},\pgfkeysvalueof{/pgfplots/ymin})},
            anchor=north,
            yshift={-\pgfkeysvalueof{/pgfplots/major tick length} + 4pt},
        },
    }
    \begin{axis}[bar shift=-16pt, nodes near coords style={xshift=0pt},
        legend pos = outer north east, legend style = {name = minimd}]
        \addplot [fill=force_color_cpu, nodes near coords=A] table [x=arch, y=minimd_ref] {single_cpu_force.txt};
        \addplot [fill=neigh_color_cpu] table [x=arch, y=minimd_ref] {single_cpu_neigh.txt};
        \addplot [fill=other_color_cpu] table [x=arch, y=minimd_ref] {single_cpu_other.txt};
        \legend{Force, Neigh, Other}
        \addlegendimage{empty legend}
        \addlegendentry{\hspace{-.325cm}\textbf{A:} miniMD (ref)}
        \addlegendimage{empty legend}
        \addlegendentry{\hspace{-.325cm}\textbf{B:} miniMD (Kokkos)}
        \addlegendimage{empty legend}
        \addlegendentry{\hspace{-.325cm}\textbf{C:} tinyMD (AoS)}
        \addlegendimage{empty legend}
        \addlegendentry{\hspace{-.325cm}\textbf{D:} tinyMD (SoA)}
        \addlegendimage{empty legend}
        \addlegendentry{\hspace{-.325cm}\textbf{E:} tinyMD (AoSoA)}
    \end{axis}
    \begin{axis}[bar shift= -8pt, nodes near coords style={xshift=0pt},
        legend style = {at = {([yshift = -1mm]minimd.south west)},
        anchor = north west}]
        \addplot [fill=force_color_cpu, nodes near coords=B] table [x=arch, y=minimd_kokkos] {single_cpu_force.txt};
        \addplot [fill=neigh_color_cpu] table [x=arch, y=minimd_kokkos] {single_cpu_neigh.txt};
        \addplot [fill=other_color_cpu] table [x=arch, y=minimd_kokkos] {single_cpu_other.txt};
    \end{axis}
    \begin{axis}[bar shift= 0pt, nodes near coords style={xshift=0pt},
        legend style = {at = {([yshift = -1mm]minimd.south west)},
        anchor = north west}]
        \addplot [fill=force_color_cpu, nodes near coords=C] table [x=arch, y=tinymd_aos] {single_cpu_force.txt};
        \addplot [fill=neigh_color_cpu] table [x=arch, y=tinymd_aos] {single_cpu_neigh.txt};
        \addplot [fill=other_color_cpu] table [x=arch, y=tinymd_aos] {single_cpu_other.txt};
    \end{axis}
    \begin{axis}[bar shift= 8pt, nodes near coords style={xshift=0pt},
        legend style = {at = {([yshift = -1mm]minimd.south west)},
        anchor = north west}]
        \addplot [fill=force_color_cpu, nodes near coords=D] table [x=arch, y=tinymd_soa] {single_cpu_force.txt};
        \addplot [fill=neigh_color_cpu] table [x=arch, y=tinymd_soa] {single_cpu_neigh.txt};
        \addplot [fill=other_color_cpu] table [x=arch, y=tinymd_soa] {single_cpu_other.txt};
    \end{axis}
    \begin{axis}[bar shift= 16pt, nodes near coords style={xshift=0pt},
        legend style = {at = {([yshift = -1mm]minimd.south west)},
        anchor = north west}]
        \addplot [fill=force_color_cpu, nodes near coords=E] table [x=arch, y=tinymd_aosoa] {single_cpu_force.txt};
        \addplot [fill=neigh_color_cpu] table [x=arch, y=tinymd_aosoa] {single_cpu_neigh.txt};
        \addplot [fill=other_color_cpu] table [x=arch, y=tinymd_aosoa] {single_cpu_other.txt};
    \end{axis}
\end{tikzpicture}
\vspace{-3ex}
\caption{Execution time in seconds (lower is better) for force calculation and neighbor list creation a 100 time-steps simulation on CPU architectures. Simulations were performed with $32^3$ unit cells with 4~particles per unit cell. Tests were performed in a single core (no parallelism) with fixed frequency. For AVX (Broadwell), AnyDSL emitted code with worse performance due to data gather and scatter operations, therefore results for scalar instructions are shown.}
\vspace{-2ex}
\centering
\label{fig:cpu_single_node_results}
\end{figure}
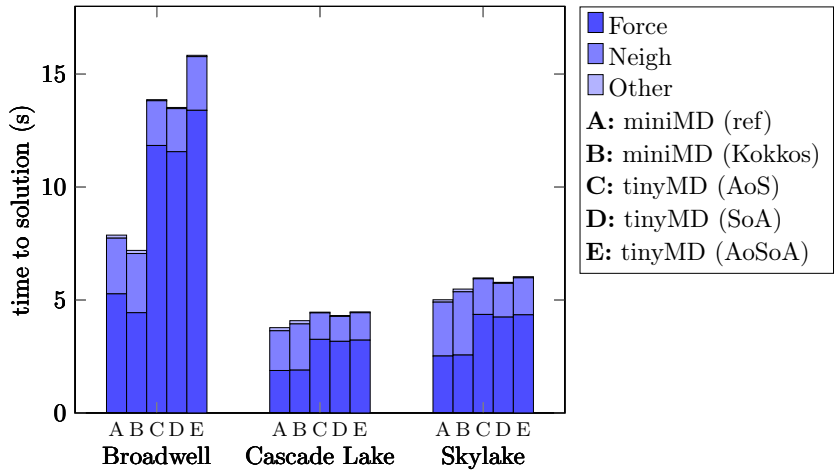

On GPU benchmarks we simulate a system configuration of $80^3$ unit cells with 4 particles per unit cell ($2.048.000$ particles in total). Both versions use the CUDA compilation tools V9.2.148, with CUDA driver version 10.2 and NVRTC version 9.1, for miniMD the Kokkos variant is required to execute on GPUs.

\colorlet{force_color_gpu}{green!60!black}
\colorlet{neigh_color_gpu}{green!40!black}
\colorlet{other_color_gpu}{green!20!black}

\begin{figure}[t]
\centering
\begin{tikzpicture}[scale=0.95]
    \pgfplotsset{
        ybar stacked, ymin=0, ymax=5, xmin=0.5, xmax=3.5, xtick=data,
        xtick={1,...,3},
        xticklabels={Pascal, Turing, Volta},
        xticklabel style={yshift=-10pt},
        ylabel={time to solution (s)}, ylabel style={yshift=-1ex},
        legend cell align={left},
        /pgf/bar width=8pt,% bar width
        scatter/position=absolute,
        node near coords style={
            font=\footnotesize,
            at={(axis cs:\pgfkeysvalueof{/data point/x},\pgfkeysvalueof{/pgfplots/ymin})},
            anchor=north,
            yshift={-\pgfkeysvalueof{/pgfplots/major tick length} + 4pt},
        },
    }
    \begin{axis}[bar shift=-12pt, nodes near coords style={xshift=0pt},
        legend pos = outer north east, legend style = {name = minimd}]
        \addplot [fill=force_color_gpu, nodes near coords=A] table [x=arch, y=minimd] {single_gpu_force.txt};
        \addplot [fill=neigh_color_gpu] table [x=arch, y=minimd] {single_gpu_neigh.txt};
        \addplot [fill=other_color_gpu] table [x=arch, y=minimd] {single_gpu_other.txt};
        \legend{Force, Neigh, Other}
        \addlegendimage{empty legend}
        \addlegendentry{\hspace{-.325cm}\textbf{A:} miniMD (Kokkos)}
        \addlegendimage{empty legend}
        \addlegendentry{\hspace{-.325cm}\textbf{B:} tinyMD (AoS)}
        \addlegendimage{empty legend}
        \addlegendentry{\hspace{-.325cm}\textbf{C:} tinyMD (SoA)}
        \addlegendimage{empty legend}
        \addlegendentry{\hspace{-.325cm}\textbf{D:} tinyMD (AoSoA)}
    \end{axis}
    \begin{axis}[bar shift= -4pt, nodes near coords style={xshift=0pt},
        legend style = {at = {([yshift = -1mm]minimd.south west)},
        anchor = north west}]
        \addplot [fill=force_color_gpu, nodes near coords=B] table [x=arch, y=tinymd_aos] {single_gpu_force.txt};
        \addplot [fill=neigh_color_gpu] table [x=arch, y=tinymd_aos] {single_gpu_neigh.txt};
        \addplot [fill=other_color_gpu] table [x=arch, y=tinymd_aos] {single_gpu_other.txt};
    \end{axis}
    \begin{axis}[bar shift= 4pt, nodes near coords style={xshift=0pt},
        legend style = {at = {([yshift = -1mm]minimd.south west)},
        anchor = north west}]
        \addplot [fill=force_color_gpu, nodes near coords=C] table [x=arch, y=tinymd_soa] {single_gpu_force.txt};
        \addplot [fill=neigh_color_gpu] table [x=arch, y=tinymd_soa] {single_gpu_neigh.txt};
        \addplot [fill=other_color_gpu] table [x=arch, y=tinymd_soa] {single_gpu_other.txt};
    \end{axis}
    \begin{axis}[bar shift= 12pt, nodes near coords style={xshift=0pt},
        legend style = {at = {([yshift = -1mm]minimd.south west)},
        anchor = north west}]
        \addplot [fill=force_color_gpu, nodes near coords=D] table [x=arch, y=tinymd_aosoa] {single_gpu_force.txt};
        \addplot [fill=neigh_color_gpu] table [x=arch, y=tinymd_aosoa] {single_gpu_neigh.txt};
        \addplot [fill=other_color_gpu] table [x=arch, y=tinymd_aosoa] {single_gpu_other.txt};
    \end{axis}
\end{tikzpicture}
\vspace{-3ex}
\caption{Execution time in seconds (lower is better) for force calculation and neighbor list creation a 100 time-steps simulation on GPU architectures. Simulations were performed with $80^3$ unit cells with 4~particles per unit cell.}
\vspace{-2ex}
\centering
\label{fig:single_gpu_results}
\end{figure}
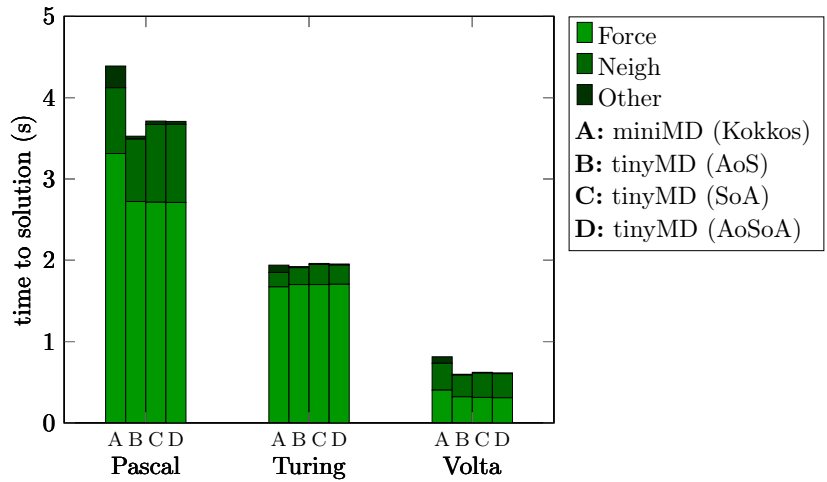

\autoref{fig:cpu_single_node_results} depicts the force calculation and neighbor list creation times for tinyMD and miniMD for the CPU architectures.
For AVX Broadwell architecture, tinyMD produced poor vectorized code because of the gather and scatter operations, and therefore the results for scalar operations are used, that is why the miniMD performance is much better for this architecture.
Note that this pairwise interactions kernels do not provide a straightforward way for vectorization, hence the compiler requires sophisticated vectorization analysis algorithms to achieve good performance within CPU cores.
For the AVX512 processors Cascade Lake and Skylake, AnyDSL produced more competitive code compared to miniMD, although the generated code is still inferior than the one generated by the Intel compiler.

This demonstrates a limitation for AnyDSL, as the auto-vectorizer is not capable of generating the most efficient variant for the potential kernels, this can be due either because of (a) the compiler itself (AnyDSL/Impala code must be compiled with Clang) or (b) the transformations performed by the AnyDSL compiler.

In all of the cases, however, it is possible to notice that the neighbor list creation performance is better for tinyMD (on AVX512 processors, it outperforms miniMD by more than a factor of two).
This can be a result of both (a) specialized code generation for both half- and full-neighbor lists creation and (b) usage of vectorization instructions to check for particle distances.
As for the data layout, the structure of arrays is the best layout for the particle \lst{Vector3D} data on CPU.
This data layout enhances locality for data in the same dimension, and therefore can improve the speedup to gather data into the vector registers.

\autoref{fig:single_gpu_results} depicts the force calculation and neighbor list creation times for tinyMD and miniMD for the GPU architectures.
For Pascal and Volta architectures, all tinyMD variants outperform miniMD. It is also noticeable that the performance advantage comes from the force compute time, which demonstrates tinyMD can generate performance-portable MD kernels to different devices.
For Turing architecture, the performance for miniMD is slightly faster than tinyMD slower variants, and the fastest one (array of structures) moderately outperforms miniMD.
The time difference for distinct \lst{Vector3D} arrays data layout on GPU is mostly concentrated in the neighbor list creation, where we can observe the array of structures delivers the best performance.

\subsection{Weak Scalability}

For the CPU weak scalability tests we chose the same miniMD setup used for the single core evaluation.
For every node involved a $96^3$ system of unit cells was included into the simulation domain. Hence, the total number of particles simulated is $3.538.944 \times \text{number\_of\_nodes}$.

The tests were executed on the SuperMUC-NG supercomputer. MPI ranks are mapped to the cores of two Intel Skylake Xeon Platinum 8174 processors (24 physical cores each) on each node. The processors are accompanied by a total of \SI{96}{GB} of main memory. All physical cores were used resulting in 48 MPI ranks per node.

\autoref{fig:weak_scaling_cpu} depicts the time to solution for both tinyMD and miniMD executing on different number of nodes. Although miniMD delivered superior results on our single core experiments, we can notice that for this configuration tinyMD presented a better performance. This happened because for this configuration the neighbor lists creation and communication times overcame the faster force field kernel in miniMD. tinyMD keeps perfect scaling for all amount of nodes, whereas miniMD starts to degrade its parallel efficiency over 512 nodes. Therefore tinyMD provides very competitive weak-scaling results in comparison to state-of-the-art applications.

\begin{figure}[t]
\centering
\begin{tikzpicture}[scale=0.80]
    \tikzstyle{every node}=[font=\small]
    \begin{axis}[
            width=\textwidth, height=8cm,
            xmode=log,log basis x={2},
            xmin=1,   xmax=2048,
            ymin=0,   ymax=5,
            log ticks with fixed point,
            xlabel={\# nodes}, xlabel style={yshift= 1ex},
            xticklabel={ % hack for precision issues with 1024 and 2048
                \pgfkeys{/pgf/fpu=true}
                \pgfmathparse{int(2^\tick)}
                \pgfmathprintnumber[fixed]{\pgfmathresult}
            },
            ylabel={time to solution (s)}, ylabel style={yshift=-1ex},
            grid=major,
            legend pos=south east,
        ]
        \addplot table[x=nodes,y=tinymd] {scaling_cpu.txt};
        \addplot table[x=nodes,y=minimd] {scaling_cpu.txt};
        \legend{tinyMD, miniMD}
    \end{axis}
\end{tikzpicture}
\vspace{-3ex}
\caption{Weak-scaling comparison between tinyMD and miniMD for CPU on SuperMUC-NG. For 256 and 1024 nodes miniMD crashed with memory violation errors and, hence, results were interpolated.}
\vspace{-2ex}
\label{fig:weak_scaling_cpu}
\end{figure}
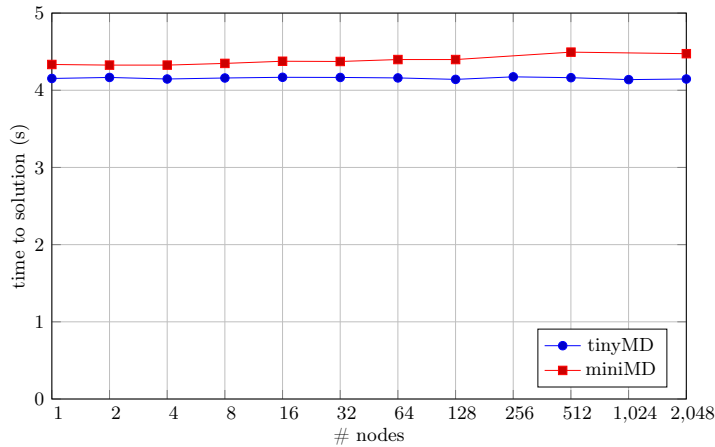

\begin{figure}[t]
\centering
\begin{tikzpicture}[scale=0.80]
    \tikzstyle{every node}=[font=\small]
    \begin{axis}[
            width=\textwidth, height=8cm,
            xmode=log,log basis x={2},
            xmin=1,   xmax=1024,
            ymin=0,   ymax=1,
            log ticks with fixed point,
            xlabel={\# nodes}, xlabel style={yshift= 1ex},
            xticklabel={ % hack for precision issues with 1024 and 2048
                \pgfkeys{/pgf/fpu=true}
                \pgfmathparse{int(2^\tick)}
                \pgfmathprintnumber[fixed]{\pgfmathresult}
            },
            ylabel={time to solution (s)}, ylabel style={yshift=-1ex},
            grid=major,
            legend pos=south east,
        ]
        \addplot table[x=nodes,y=tinymd] {scaling_gpu.txt};
        \legend{tinyMD}
    \end{axis}
\end{tikzpicture}
\vspace{-3ex}
\caption{Weak-scaling results for tinyMD on the Piz~Daint supercomputer. Each process is mapped to a node with one NVIDIA Tesla P100 16GB GPU.}
\vspace{-2ex}
\label{fig:weak_scaling_gpu}
\end{figure}
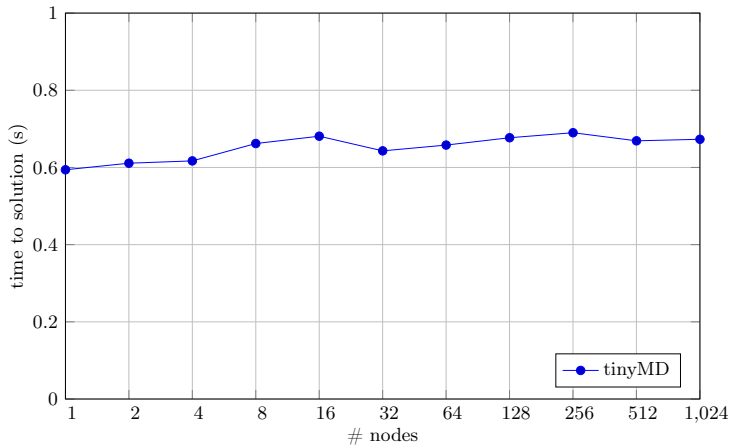

We performed the experimental tests for GPU weak-scalability on the Piz~Daint supercomputer using the XC50 compute nodes.
Each node consists of a NVIDIA Tesla P100 16GB GPU, together with an Intel Xeon E5-2690 v3 @ 2.60GHz processor with 12 cores, and \SI{64}{GB} of main memory.
Each MPI rank is mapped to a GPU in our simulation---so one rank per node.
For each GPU, a $50^3$ system of unit cells was included in the simulation domain, resulting in a total of $500.000 \times \text{number\_of\_gpus}$ particles.

\autoref{fig:weak_scaling_gpu} depicts the time to solution for tinyMD on GPU. From 1 to 4 nodes, tinyMD executes faster than for more nodes, which is expected because with less than 8 nodes, there is no remote communication in all directions.
Since GPU compute kernels are much faster compared to CPU, remote communication consumes a bigger fraction of the total time and this can affect the scalability.
From 8 to 1024 nodes where remote communication is performed in all directions, it is reasonable to state that tinyMD presented perfect scalability.

For miniMD, we were not able to produce weak-scalability results because we could not manage to compile it properly on the Piz~Daint GPU cluster. The current compiler versions available in the cluster delivered error messages during compilation and the ones that were able to compile generated faulted versions of miniMD, which delivered unclear MPI errors during runtime. Nevertheless, the presented results for tinyMD are enough to demonstrate its weak-scaling capability on GPU clusters.

\subsection{Load Balancing}

For the load balancing experiments, we also execute our tests on SuperMUC-NG with 48 CPU cores per node. In this experiments, the Spring-Dashpot contact model was used, with the stiffness and damping values set to zero, which means that particles keep static during the simulation. In order to measure the balancing efficiency, we distribute the particles evenly in half of the domain, using a diagonal axis to separate the regions with and without particles. When more nodes are used in the simulation, the domain is also extended in the same proportion, hence the proportion of particles to nodes remains the same.

We perform 1000 time steps during the simulation, and the load balancing is performed before the simulation begins. This performs a good way to measure the load balancing efficiency because we can expect an improved speedup close to a factor of two. We use a system of $96^3$ unit cells per node (48 cores), with $442.368$ particles per node. We performed experiments with Hilbert and Morton space-filling curves methods to balance the simulation.

\begin{figure}[t]
\centering
\begin{tikzpicture}[scale=0.80]
    \tikzstyle{every node}=[font=\small]
    \begin{axis}[
            width=\textwidth, height=8cm,
            xmode=log,log basis x={2},
            xmin=1,   xmax=2048,
            ymin=0,   ymax=45,
            log ticks with fixed point,
            xlabel={\# nodes}, xlabel style={yshift= 1ex},
            xticklabel={ % hack for precision issues with 1024 and 2048
                \pgfkeys{/pgf/fpu=true}
                \pgfmathparse{int(2^\tick)}
                \pgfmathprintnumber[fixed]{\pgfmathresult}
            },
            ylabel={time to solution (s)}, ylabel style={yshift=-1ex},
            grid=major,
            %legend pos=north east,
            legend pos=south east,
        ]
        \addplot table[x=nodes,y=morton]  {lb_cpu.txt};
        \addplot table[x=nodes,y=hilbert] {lb_cpu.txt};
        \addplot table[x=nodes,y=no_lb]   {lb_cpu.txt};
        \legend{Morton, Hilbert, Imbalanced}
    \end{axis}
\end{tikzpicture}
\vspace{-3ex}
\caption{Load-balancing results for tinyMD on SuperMUC-NG with the Spring-Dashpot contact model. For each node involved, the domain is extended by a $96^3$ system of unit cells, and particles are then distributed through half of the domain using a diagonal axis.}
\vspace{-2ex}
\label{fig:load_balancing}
\end{figure}
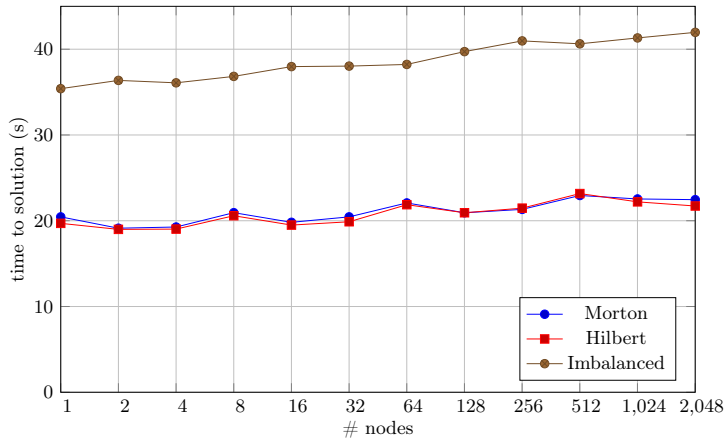

\autoref{fig:load_balancing} depicts the time to solution for the load balancing experiments. Both balanced and imbalanced simulations scale well during the experiments, and it is also possible to notice the performance benefit from the balanced simulations. The benefit as previously mentioned is close to a factor of two, and the difference between Morton and Hilbert methods is not too significant, with Hilbert being more efficient at some node sizes.

Although the load balancing feature works on GPU simulations, the communication code using waLBerla domain partitioning takes a considerable fraction of their time. Therefore a different strategy on this communication code is necessary for GPU accelerators to reduce the fraction of the communication time compared to the potential kernels and then achieve benefits from load-balancing. For this reason, we just show the experimental results on SuperMUC-NG for CPU compute nodes.

\section{Conclusion}
\label{sec:concl}

This paper presents tinyMD: an efficient, portable, and scalable implementation of an \ac{MD} application using the AnyDSL partial evaluation framework.
To evaluate tinyMD, we compare it with miniMD, implemented in C++ that relies on the Kokkos library to be portable to GPU accelerators.
We discuss the implementation differences regarding code portability, data layout and MPI communication.

To achieve performance-portability on most recent processors and supercomputers, we provide abstractions in AnyDSL that allow our application to be mapped to distinct hardware and be compiled with different data layouts.
All this can be done with zero overhead due to partial evaluation, which is one of the main advantages we get when using AnyDSL.

Moreover, we also couple our application with the waLBerla multi-physics framework to use its load balancing mechanism within tinyMD simulations.
This emphasizes how our Impala code can be coupled to other implementations, avoiding doing work that otherwise would not generate benefits.
Furthermore, we show how this coupling can be made easier when using higher-order functions in tinyMD communication code to abstract the communication pattern.
This permit us to insert the waLBerla communication logic into tinyMD by just replacing functionality.

Performance results for tinyMD on single CPU and GPU show that it is competitive with miniMD on both single CPU cores and single GPU accelerators, as well as on supercomputer running on several compute nodes.
Weak scalability results on the SuperMUC-NG and Piz~Daint supercomputers demonstrate that tinyMD achieves perfect scaling on top supercomputers for the presented benchmarks.
The load-balancing results on SuperMUC-NG demonstrate that our strategy for coupling tinyMD and waLBerla works as expected, since the balanced simulations in our experiments reach a speedup close to a factor of two compared to the imbalanced simulations when filling just half of the domain.

\bibliography{tinyMD_paper}

\end{document}